\documentstyle[aps,epsf,twocolumn,prb]{revtex}
\begin{document}
\def\ef{\epsilon_{\rm F}}
\def\efb{\epsilon_F^{>}}
\def\rof{\rho_0(\epsilon_{\rm F})}
\twocolumn[\hsize\textwidth\columnwidth\hsize\csname@twocolumnfalse\endcsname

\title{Heavy-fermion and spin-liquid behavior
in a Kondo lattice with magnetic frustration}
\author{S. Burdin${}^{1,2}$, D. R. Grempel${}^{3}$ and A. Georges${}^{4}$
}
\address{$^1$D\'epartement de Recherche Fondamentale sur la Mati\`ere
Condens\'ee,\\ SPSMS,  CEA-Grenoble,
38054 Grenoble Cedex 9,
France.\\
${}^{2}$ Institut Laue-Langevin, B.P. 156, 38042 Grenoble Cedex 9, France.\\
${}^{3}$ CEA-SACLAY, SPCSI, 91191
Gif-sur-Yvette Cedex, France.\\
${}^{4}$CNRS - Laboratoire de Physique Th{\'e}orique, Ecole Normale
Sup{\'e}rieure, 24 Rue Lhomond 75005 Paris, France.}
\date{\today}
\maketitle
\widetext
\begin{abstract}
\noindent
We study the competition between the Kondo effect and frustrating
exchange interactions in a Kondo-lattice model within a
large-${\cal N}$ dynamical mean-field theory. We find
 a $T=0$ phase transition between a heavy Fermi-liquid and a
spin-liquid for a critical value of the exchange $J_c =
T_{K}^0$, the single-impurity Kondo temperature. Close to the
critical point, the Fermi liquid coherence scale $T^\star$ is
strongly reduced and the effective mass strongly enhanced. The
regime $T>T^\star$ is characterized by spin-liquid magnetic
correlations and non-Fermi-liquid properties. 
It is suggested that magnetic frustration is a general mechanism 
which is essential to explain the large effective mass of 
some metallic compounds such as  LiV$_2$O$_4$. 
\end{abstract}
]
\narrowtext

\section{Introduction}
The interplay between the Kondo effect and RKKY interactions
is an essential feature of heavy-fermion systems. In dilute Kondo
systems local magnetic moments are screened
 below the single-site Kondo temperature
$T^0_K$ and a local Fermi-liquid (FL) picture
applies~\cite{Hewson}. In dense systems, intersite magnetic
interactions compete with the Kondo effect, leading to a  quantum
phase transition at which the metallic paramagnetic ground state
becomes unstable~\cite{Doniach} when their strength attains a critical value
$J_c$. In the vicinity of the quantum critical
point (QCP) the physical properties of a large class of strongly
correlated metals differ strikingly from those of normal Fermi
liquids~\cite{nfl-exp}.

The origin of non-Fermi liquid (NFL)
 behavior in the quantum critical region of heavy fermion systems
is an issue of current theoretical interest~\cite{catherine}. Two
scenarios have been proposed for the case of the antiferromagnetic
QCPs~\cite{qimiao}. In the first one, 
Kondo screening of the local moments takes place below a
  temperature $T_K$ that stays finite throughout the
 paramagnetic phase (including the QCP).  At
 a lower coherence temperature $T^{\star}$, a heavy Fermi
 liquid of composite quasi-particles forms  and the magnetic phase transition
 at $J_c$ is driven by a spin-density-wave (SDW)
 instability of the Fermi surface. NFL behavior around the
 QCP results from the coupling of the heavy electrons
to critical long-wavelength SDW fluctuations~\cite{millis-hertz}.

In the second scenario, intersite interactions  are strong enough to prevent 
Kondo screening from occurring at the critical
coupling. Both $T_K$ and $T^{\star}$ are expected to vanish  at
$J_c$, leading to the dissociation of the composite heavy
quasi-particles into decoupled local moments and conduction
electrons. In this case, NFL behavior is a consequence of the
critical properties of the local spin fluctuations that are
associated to the process of Kondo screening.

In systems with magnetic frustration, due either to the geometry
of the lattice\cite{ramirez} or to disorder, conventional magnetic
ordering may give way to spin-glass (SG) freezing, or be
suppressed altogether, leaving a correlated paramagnet or
``spin-liquid'' (SL) state.

In this paper we consider a Kondo-lattice model
with frustrated magnetic interactions between the localized spins. 
We solve this model using a combination of dynamical mean-field theory and 
large-N techniques.
We find a quantum critical point (QCP) between a Fermi
liquid (FL) and a SL phase at a critical coupling $J_c = T^0_K$,
the single-impurity Kondo temperature. We show that near the QCP
the coupling of conduction electrons to local critical
spin-fluctuations leads to the suppression of both the Kondo scale
$T_K$ and of the FL coherence scale $T^\star$. In addition,  
$T_K/T^{\star} \gg 1$ for $J \sim J_c$, and the effective mass is drastically 
enhanced by the combined effect of the frustration and the Kondo effect 
near the QCP.
This is reminiscent of the second scenario described above. However, it should 
be emphasized that the strong quantum fluctuations associated with the (fermionic) 
large-N limit considered here prevent any type of magnetic ordering to take place. 
As argued in the conclusion of the paper, corections beyond that limit will 
reintroduce (spin-glass) long-range order. At least in high enough dimensions, this will 
usually happen at a {\it smaller} value of the magnetic coupling $J$ than 
the one corresponding to the vanishing of the coherence scale. Nevertheless, 
the existence of a very small coherence scale and large effctive mass near the 
QCP, as well as that of an intermediate non-FL crossover regime between 
for $T^*<T<T_K$ are robust features.     
\begin{figure}
\epsfxsize=2in
\epsffile{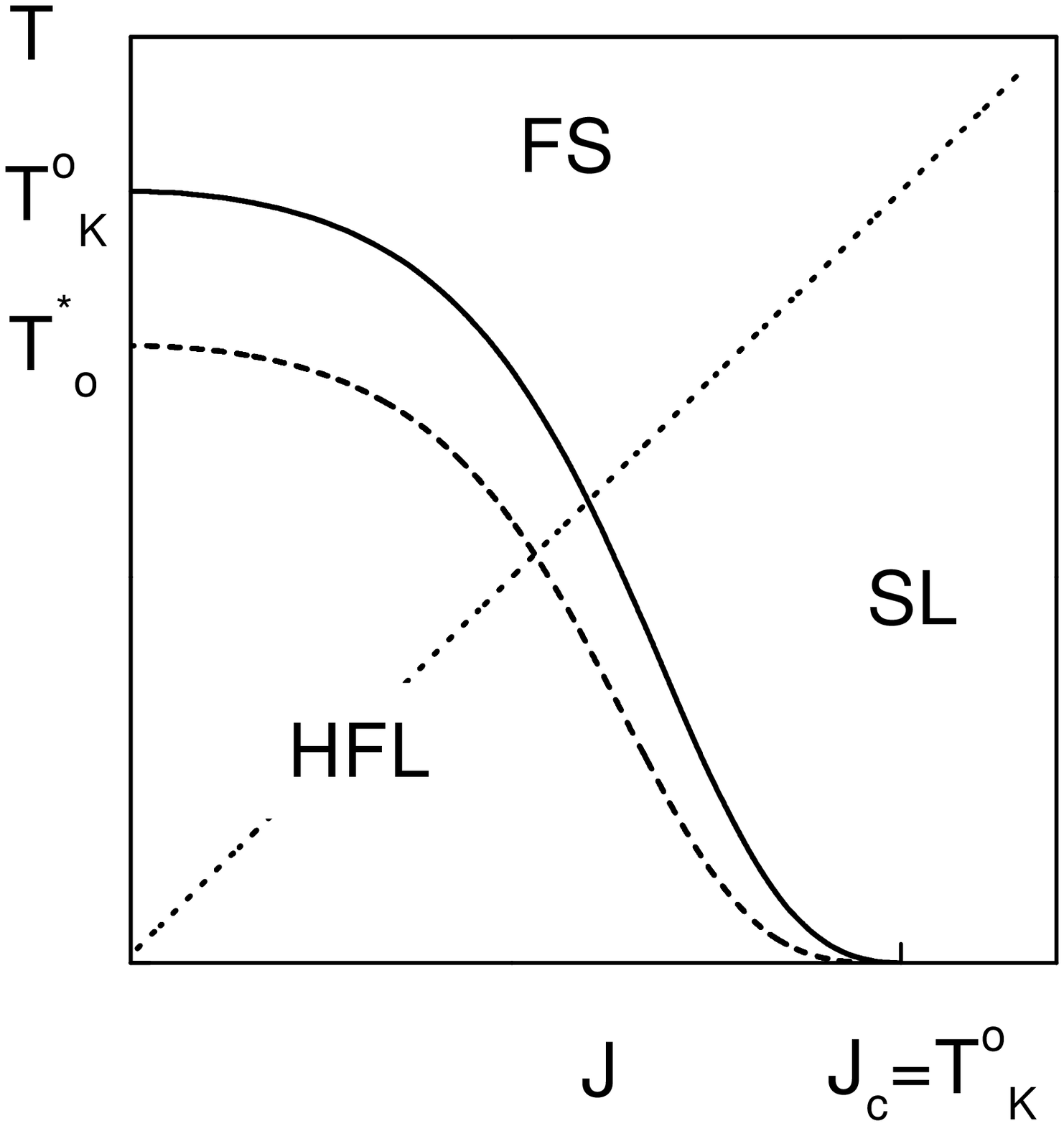}
\caption{Schematic phase
diagram of the model in the $J - T$-plane. Kondo temperature (solid
line) and
 coherence  temperature
(dashed line) as functions of $J$ for fixed values of $J_{ K}$ and
$n_{c}$.
The system is a heavy Fermi liquid below $T^{\star}(J)$. Above the line
$T_K(J)$,
the localized spins are essentially free (FS) for $J < T$,  while
they form a highly correlated
spin liquid for $J > T$. Near $J_c$ the spin liquid correlations
start building up just above $T^{\star}$. All the lines represent crossovers.} \label{f.1}
\end{figure}
Below $T^{\star}$ the low temperature properties of the system may
be described in terms of heavy quasi-particles whose mass
$m^{\star}$ diverges at the QCP.  In the regime $T^{\star} < T <
T_K$, the electrons form an incoherent ``bad'' metal. They
gradually decouple from the localized spins for $T>T_K$, while the
local spin dynamics remains SL-like for $T<J$. In the whole range
$T^\star<T<J$, the spin and transport properties are markedly
different from those of a FL. This type of non-Fermi liquid regime
has been previously studied in Ref.~\onlinecite{page} in the
different context of a doped Mott insulator. As we shall see
below, the physical properties of our model share common features
with the experimentally observed behavior of LiV$_2$O$_4$, the
first compound which displays heavy-fermion behavior even though
no f-electrons are involved~\cite{exp1,kuyomi,exp2}.

\section{Model}
We study the Kondo-lattice model defined by the Hamiltonian :
\begin{equation}
\label{hamil}
 H=-\sum_{i,j,s}\; t_{ij}\; c_{i s}^{\dagger}c_{j s}^{}
  + J_{ K} \sum_{i} \vec{S}_i
\cdot \vec{s}_{i} + \sum_{<i,j>} J_{ij}\; \vec{S}_{i}
 \cdot \vec{S}_{j} \; ,
\end{equation}
where $\vec{S}_i$ and $c_{i s}^{\dagger}$ represent respectively a
localized spin and a conduction-electron creation operator at the
$i-$th site of the lattice. The localized spins interact with the
conduction-electron spin density $\vec{s}_{i}=1/2\sum_{s,s'}\;c_{i
s}^{\dagger} \vec{\sigma}_{s s'} c_{i s'}^{}$ {\it via} a local
Kondo coupling $J_{ K}$ and $J_{ij}$ is the magnetic exchange
coupling between nearest-neighbor pairs of localized spins. The
effects of frustration are introduced in the model by an
appropriate choice of the couplings $J_{ij}$. In a model for
geometrically frustrated materials these must be taken
anti-ferromagnetic on lattices of the Kagome or pyrochlore
type~\cite{ramirez}. In the case of
 a metallic SG,
the couplings can be taken as a set of random variables with zero
mean and variance $J/\sqrt{z}$ (z is the coordination of the
lattice).
 In the following, we focus on the case of random
$J_{ij}$. This should be viewed simply as a way to
generate a SL regime with non-trivial spin dynamics, and we expect
our conclusions to be similar in the case of geometric frustration.

The problem defined by Eq.~(\ref{hamil}) is tractable in the limit
of large coordination $z \to \infty$ where dynamical mean-field
theory (DMFT) is  applicable~\cite{dmft} and the model can be
reduced to an effective single-site theory. Using standard methods
to perform the average over the disorder~\cite{bm} the DMFT action
associated to Eq.~(\ref{hamil}) may be written in the
form~\cite{footnote1}:
\begin{eqnarray}
\label{action} \nonumber {\cal A}&=&\sum_{s}\int_0^\beta d\tau
\int_0^\beta d\tau'\;c^{\dagger}_s(\tau)\;\left[
\left(\;\partial_\tau  - \mu \;\right)\;\delta(\tau - \tau')
\right. \\
& & \left.   - \;t^2 \;G_c(\tau-\tau')\right]c^{}_{s}(\tau') +
\;J_K\; \int_0^\beta d\tau
\;\vec{S}(\tau) \cdot \vec{\sigma}_c(\tau) \\
\nonumber & & -  \frac{J^2}{2}\;\; \int_0^\beta d\tau \int_0^\beta
d\tau'\;\chi(\tau-\tau')\; \vec{S}(\tau) \cdot \vec{S}(\tau')\;,
\end{eqnarray}
where $\vec{\sigma}_c =
 \sum_{s,s'}\vec{\sigma}_{s s'} c^{\dagger}_s(\tau)
c^{}_{s'}(\tau) $ is the conduction electron spin density, $\mu $
is the chemical potential and the rescaling $t \to t/\sqrt{z}$ was
performed in order to obtain finite expressions in the limit
$z\to\infty$. $G_c(\tau)$ and $\chi (\tau)$ are the
conduction-electron Green function and the localized spin
autocorrelation function, respectively. These are determined by
the self-consistency conditions~\cite{dmft,bm} $G_c(\tau) =
-\langle\;c^{}_s(\tau)\;c^{\dagger}_s(0)\; \rangle_{\cal A}$ and
$\chi (\tau)=\langle\;\vec{S}(\tau).\vec{S}(0)\;\rangle_{\cal A}$.
The first two terms in Eq.~(\ref{action}) represent the action of
the Kondo-lattice hamiltonian. The last term is the action of the
quantum SG model recently studied by several
authors~\cite{saye,grro2,gepasa}. It was shown that for $S=1/2$
the ground state has SG order ~\cite{grro2}. However, SL solutions
appear above the SG transition temperature, specially in the
formal limit of small values of $S$~\cite{saye,gepasa}.
Furthermore, it has been recently suggested \cite{gesiflo} that
the effective action describing the spin dynamics in this regime
is also relevant for geometrically frustrated antiferromagnets.

The SL solutions are well described in the large-${\cal N}$
approach that we discuss next.

\section{Large-${\cal N}$ solution}
The model defined in Eq.~(\ref{action}) cannot be solved
analytically as it stands, but much progress can be made by
solving it in the large-${\cal N}$ approach which has been
extensively used in the study of the Kondo lattice~\cite{LargeN}
and also allows to deal with the magnetic exchange term in
Eq.~(\ref{action}) \cite{saye,gepasa,page}. In this approach the
spin symmetry is extended to SU(${\cal N}$) and the coupling
constants are rescaled as $J_K \to J_K/{\cal N}$, $J \to
J/\sqrt{{\cal N}}$. The localized spin is represented in terms of
fermion operators, $S^{\sigma\sigma'} =
f^{\dagger}_{\sigma}f_{\sigma'}-\delta_{\sigma\sigma'}/2$, subject
to the local constraint
$\sum_{\sigma}f^{\dagger}_{\sigma}f_{\sigma}={\cal N}/2$.  The
interaction terms in Eq.~(\ref{action}) become now quartic. These
are decoupled introducing Hubbard-Stratonovich fields $B(\tau)$
(conjugate to $\sum_{\sigma}f_{\sigma}^{\dagger}c_{\sigma}$) and
$P(\tau,\tau')$ (conjugate to
$\sum_{\sigma}f_{\sigma}^{\dagger}f_{\sigma}$) and the constraint
is enforced through the introduction of a Lagrange multiplier $i
\lambda(\tau)$. In the ${\cal N} \to \infty$ limit the physics is
controlled by a saddle point at which the Bose field condenses
$\langle B(\tau)\rangle_{\cal A} =r$, and the Lagrange multiplier
takes a static value $i\lambda(\tau)=\lambda$, while
$P(\tau,\tau')=P(\tau-\tau')$ generates a frequency dependent
local self-energy \cite{saye,page}. The saddle point equations can
be written in the following compact form:
\begin{eqnarray}
\label{saddle}
\left\{ \begin{array}{c} -r/J_{
K}\\
1/2\\
n_{c}/2 \end{array} \right\} = -\frac{1}{\pi}
\int_{-\infty}^{\infty}\;d\omega\;n_{F}(\omega) \; {\rm
Im}\;\left\{
\begin{array}{c}
G_{fc}(\omega)\\ G_{f}(\omega) \\ G_{c}(\omega) \end{array}
\right\} \; ,
\end{eqnarray}
where $n_{F}$ is the Fermi function and $G_{c}$, $G_{f}$ and
$G_{fc}$ are the full conduction-electron, f-electron and mixed
Green functions, respectively given by :
\begin{eqnarray}
\nonumber G_{c\;}(\omega) &=& G_{c}^{0}(\omega+\mu -r^2{\cal G}_{f}(\omega)) \\
\label{Green} G_{f}(\omega) &=& {\cal G}_{f}(\omega)\;\left[\; 1 + r^2\;{\cal
G}_f(\omega)\;G_{c}(\omega)\;\right]\;\\
\nonumber G_{fc}(\omega) &=& r\;{\cal G}_{f}(\omega)\;G_{c}(\omega)
\;.
\end{eqnarray}

Here, $G_{c}^{0}(\omega)=\sum_{k}1/(\omega-\epsilon_{k})$ is the
non-interacting electronic local Green function, and we introduced
the ``bare'' f-electron Green function
\begin{equation}
\label{bare-f}
{\cal
G}_{f}(\omega) = {1 \over \omega+\lambda-\Sigma_{loc}(\omega)}\;,
\end{equation}
where the local self-energy is 
\begin{equation}
\label{sigma} \Sigma_{loc}(\tau)=-J^{2}\;G_{f}^{2}(\tau)\;
G_{f}(-\tau) \; .
\end{equation}

\subsection{The Kondo temperature}

At high temperature or for large values of $J$ the only solution
of Eqs.~(\ref{saddle})-(\ref{sigma}) has $r=\lambda=0$. This
represents a regime in which the localized spins and the
conduction electrons are decoupled. In this regime ${\cal
G}_f(\omega)\;=\;G_S(\omega)$, the solution of a non-linear
integral equation first investigated by Sachdev and Ye
~\cite{saye} that reads:
\begin{eqnarray}
\label{sachdev_gf} G_{S}(i\omega_n)& = &
\left[\;i\omega_n - \Sigma_{S}(i\omega_n)\;\right]^{-1}\\
\label{sachdev_sigma} \Sigma_{S}(\tau) &=& -J^2 
\left[\;G_{S}(\tau)\;\right]^2 \;G_{S}(-\tau)\;.
\end{eqnarray}
In the region ${\rm
max}(T,\omega) < J$ 
the solution of this solution describes a SL with a non-trivial
local spin dynamics characterized by a slow decay of the local spin
autocorrelation function~\cite{saye}, 
$\langle \vec{S}(0)\cdot\vec{S}(t)\rangle \sim 1/t$.

In the large-${\cal N}$ theory, the onset of Kondo screening is
signaled by a phase transition at a critical temperature $T_{ K}$
at which a second solution with $r\not= 0$ appears. The equation
for $T_{ K}$ is thus
\begin{equation}
\label{Tk} \frac{1}{J_{K}}=\int_{-\infty}^\infty
\;\frac{d\omega}{\pi}\;n_{F}(\omega)\; {\rm Im}\;\left[\;
{G_{c}^{0}(\mu +\omega) G_{S}(\omega)}\;\right]\;,
\end{equation}
where $\mu$ is the chemical potential for
free conduction electrons with density $n_{ c}$ at $T=T_{K}$.  
In the limit $J \rho_{0}(\mu) \ll 1$
 this equation can be cast in the form
\begin{eqnarray}
\nonumber \frac{1}{J_{\sc k}} & = & -
 \frac{\rho_0(\mu)}{2}\;\int_{-\infty}^{\infty} d\epsilon\;\left(\;{G'}_{S}(\epsilon) - \frac{1}{\epsilon}\;\right)\; 
\tanh\left(\frac{ \epsilon}{2 T_K}\right)\\
\label{eq_for_tk} & + &  \frac{1}{2} \int_{-\infty}^{\infty} 
\frac{d\epsilon}{\epsilon}\; \rho_0(\mu +
\epsilon)\;\tanh\left(\frac{ \epsilon}{2 T_K}\right)\;,
\end{eqnarray}
where ${G'}_{S}$ denotes the real part of the f-electron Green function
and $\rho_{0}$ is  the conduction electron density of
states (d.o.s).

For $J \to 0$, $G_S(\omega) =
\omega^{-1}$ and only the second integral on the right-hand side of 
Eq.~(\ref{eq_for_tk}) survives. 
Then we find that $T_K=T_K^0$, the single-impurity
Kondo scale, given  in the weak-coupling limit $J_{K}\rho_{0}(\epsilon_F) \ll
1$ by~\cite{bugegr}:
\begin{equation}
\label{t0k} T^0_K=D\;e^{-1/ (J_{K}\rho_{0}(\epsilon_F))}
\;\sqrt{1-(\epsilon_{F}/D)^2}\;F_{K}(n_{c})
\end{equation}
where $\epsilon_{F}$ is the non-interacting Fermi level
 and
\begin{equation}
\label{Fk} 
\ln F_{K}(n_c) = \int_{-(D +
\epsilon_F)}^{D - \epsilon_F}\; \frac{d\epsilon}{|\epsilon|}\; 
\frac{\;\rho_0(\epsilon_F + \epsilon) - \rho_0(\epsilon_F)\;}{2 \rho_0(\epsilon_F)}\;,
\end{equation}
depends on the details of the band structure and the filling of the
conduction band but not on the Kondo coupling~\cite{bugegr}.

For $J \ll T^0_K$ the intersite coupling is a small perturbation
and it may be shown that $T_K = T^0_K\; \left[ 1 - {\cal
O}\left(\left(J/T^0_K\right)^2\right)\right]$. In the opposite limit,
 $J\gg T_K^0$, the magnetic exchange dominates and it can be shown
that the form  of the decay
of the spin autocorrelation function in the SL phase
implies that the Kondo effect cannot
take place. As a result, for $J$ above a critical value $J_c =
T_K^0$, the conduction electrons remain decoupled from the localized spins
down to zero temperature (Fig.~1). This is expected from a comparison of
the binding energy of two localized spins in
the SL regime ($\sim J$) and the energy gained by forming singlets
between the spins and the conduction electrons which is at most
${\cal O}(T^0_K)$. 

In order to find the behavior of  $T_K$ close to
the critical point,  
we evaluate the first integral on the
 right-hand side of Eq.~(\ref{eq_for_tk}) using  
the asymptotic form~\cite{page} of $G_S(\omega)$,
\begin{eqnarray}
\label{asymptotic} 
G_{S}(\omega) & \sim & \left\{ 
\begin{array}{ccc}\frac{\Gamma\left[ 1/4 \right]^2}{2
\pi^{5/4}}\;{\omega/(2 T) - i \over \sqrt{J T}}
\;, & \left|\omega\right| < T \ll J\;,\\
\\ \left(\;\frac{\pi}{4}\;\right)^{1/4}\;{ {\rm sign}(\omega) - i
\over \sqrt{J |\omega|}}\;, & T <
\left| \omega \right| \ll J\;.
\end{array}
\right.
\end{eqnarray}
To leading order in  $T_K/J$ we find:
\begin{equation}
\label{first_int}
 I_1  \sim \rho_{0}(\mu) \left[\;
 \ln\left(T_{K}/J\right) - a \sqrt{T_{K}/J}  \;\right]\; ,
\end{equation}
where $a$ is a numerical constant of ${\cal O}(1)$. Combining
Eqs.~(\ref{eq_for_tk}),  (\ref{t0k}) and (\ref{first_int}) we find
that, for $J \sim J_c = T^0_K$, 
\begin{equation}
\label{tk}T_{K} \approx J\;\ln^2 \left(\;\frac{T^0_K}{J}\;\right)
\sim T^0_K\;\delta^2\;,
\end{equation}
where $\delta = \left( J_c - J \right)/J_c$ measures the distance
to the QCP where the Kondo temperature
vanishes quadratically. 

\subsection{The heavy Fermi-liquid regime}
When $r$ is finite, Eqs.~(\ref{saddle})-(\ref{sigma}) admit FL
solutions at low enough temperatures. We discuss first the case $T
= 0$. A calculation analogous to that performed in Ref.~\onlinecite{page}
yields the value  of the
self-energy at zero-frequency :
\begin{equation}
\label{Luttinger} r^2/\left(\;\Sigma_{loc}(0)-\lambda\;\right) =
\epsilon_{F}^{>}-\mu\; ,
\end{equation}
In this expression, $\epsilon_{F}^{>}$ is the non-interacting
Fermi level corresponding to an electron density $(n_{c}+1)/2$ per
spin component. Eq.~(\ref{Luttinger}) implies that Luttinger's theorem is
satisfied, with a ``large'' Fermi surface containing both
conduction electrons and localized spins. In the weak-coupling
limit, $\mu\approx \epsilon_F$, the non-interacting Fermi level
corresponding to a density $n_c$.
It can be shown from Eqs.~(\ref{Green}), (\ref{sigma}) and
(\ref{Luttinger}) that the f-electron d.o.s at the Fermi level
$\rho_f(0)= {\cal O}\left(\;D/r^2\;\right)$ is finite for $r \ne
0$. Since in the SL phase $\rho_f(\omega) \propto
1/\sqrt{J\;\left| \omega \right|}$, a crossover between the FL and
SL regimes is expected at a scale $T^{\star} \sim (r^2/D)^2/J$.
We note that, for $J=0$, the coherence
scale that controls all physical quantities at low temperature
 is $r^2/D$ ~\cite{bugegr}. For $J\neq 0$, this role is
played by the much smaller scale $T^{\star}$, as detailed below.

We found that, at $T=0$, in the weak-coupling limit, the
 full set of Eqs.~(\ref{saddle})-(\ref{sigma}) can be solved
analytically in the vicinity of the QCP by using the following 
{\it Ansatz} for the f-electron d.o.s :
\begin{eqnarray}
\label{Ansatz}
\rho_{f}(\omega) = \left\{
\begin{array}{lll}
(\epsilon_F^>-\epsilon_F
)^2\;\rho_{0}(\epsilon_F^>)/r^2 & \mbox{for}\;\;\omega
\le T^{\star} \\
\\
\left(\;4
\pi^3\;\right)^{-1/4}\;\left(\;J\;|\omega|\;\right)^{-1/2} 
& \mbox{for}\;\; T^{\star} < \omega < J
\end{array}\;.
\right.
\end{eqnarray}
The first line in the above equation is $\rho_f(0)$ as determined
from Eqs.~(\ref{Green}) and (\ref{Luttinger}); the second line is
the SL density of states~\cite{saye} and $T^{\star}$ is defined 
as the energy
at which the two expressions match, {\it i.e.}:
\begin{equation}
\label{matching} T^{\star} \propto \;\frac{1}{J}\;\left(\;
\frac{r^2}{\left(\epsilon_F^> - \epsilon_F\right)^2\;  
\rho_{0}(\epsilon_F^>)}\;\right)^2\;.
\end{equation}
This establishes a relationship between  $T^{\star}$ and  $r$ that is
determined  from the equation
\begin{equation}
\label{eq_for_r}
 \frac{1}{J_{K}}= \int_{-\infty}^{0} \frac{d\omega}{\pi}\;{\rm
Im}\left[\;{\cal G}_{f}(\omega)\;G^0_{c}\left(\omega - r^2
{\cal G}_{f}(\omega)\right)\;\right]\;.
\end{equation}
The bare f-Green function ${\cal G}_{f}(\omega)$ may
 be computed using Eqs.~(\ref{Green}),
(\ref{sigma}) and (\ref{Ansatz}) and the integral in
 Eq.~(\ref{eq_for_r}) may be evaluated in the limit  $J \to J_c$. We find :
\begin{equation}
\label{tstar} 
T^{\star} \sim T^0_K\;\delta^2/\left(\;\ln \delta\;\right)^2\;.
\end{equation}
$T^{\star}$ and $T_K$ thus vanish
simultaneously at the critical coupling, with $T_K/T^{\star} \sim \ln^2 \delta
\gg 1$ as $\delta \to 0$.

The various physical regimes that follow from these
considerations are depicted schematically in Fig.~\ref{f.1}. Above
the scale $T_{ K}(J)$, the conduction electrons and the localized
spins are decoupled. This is clearly an unrealistic feature of the
large-${\cal N}$-limit. For finite ${\cal N}$, the phase
transition at $T_K$ will be replaced by a gradual decoupling of
the electrons and the spins as the temperature is raised. For
$T_K<T<J$, the localized spins remain strongly correlated in the
SL state. For $T<T^\star$, a heavy Fermi liquid  with a
large Fermi surface is formed. The effective mass of quasiparticles (given
by the inverse of the quasiparticle residue $Z$) is $ m^*/m =1/Z
\sim D/T^\star \sim D/T_K^0\,(\ln\delta/\delta)^2$. This is one of the
key results obtained in this paper: it demonstrates how the
frustrating magnetic exchange leads to a dramatic enhancement of
the effective mass with respect to the
``bare'' Kondo scale. The underlying mechanism is the large
entropy of the SL state at low temperature. The intermediate
 range $T^\star<T<T_K$
corresponds to a  crossover regime in which, as the temperature
decreases,
 the electrons
gradually couple to the localized spins, and the spin
correlations change from SL-like at high temperature to FL-like at
low temperature.

\section{Physical properties}

The physical properties of the system can be computed in a standard
way from  the Green functions.
We find that, for $T \ll T^{\star}$,
 the entropy is dominated by the quasi-particle
contribution, $S \propto T\;\rho_f(0)\;(1 - \partial
\Sigma(\omega)/\partial \omega\;|_{\omega=0})$. Each of last the two
factors gives a contribution proportional to
$\left(J/T^{\star}\right)^{1/2}$. Thus, in the FL region
$C \propto T/T^{\star}$.
In the SL regime for $T^{\star} < T \ll J$ we find $C\propto
\sqrt{T/J}$. The specific heat thus has a peak at $T \sim T^{\star}$.

In the Fermi-liquid region we find the local spin susceptibility
 $\chi_{loc}''(\omega) \propto
\omega/T^{\star}$. At $T \sim T^{\star}$ there is  a crossover to
the SL form $J\chi_{loc}''(\omega) \propto \tanh
\omega/2T$~\cite{saye,page}. Hence, the NMR spin-lattice relaxation rate
$1/T_1\propto T\,\lim_{\omega\to 0}\;\chi_{loc}''(\omega)/\omega$
obeys Korringa's law $1/T_1 \sim
T/(JT^\star)$ below $T=T^{\star}$, but is $T$-independent
above this temperature, $1/T_1 \sim 1/J$. The static local susceptibility increases
logarithmically with decreasing temperature for $T > T^{\star}$,
$\chi_{loc}(T) \propto J^{-1}\;\ln(J/T)$,  saturating to a
constant value $\propto J^{-1}\;\ln(J/T^{\star})$ for
$T<T^{\star}$.
In analogy with the result found in the closely related case of the
doped Mott insulator~\cite{page} we expect a finite uniform spin
 $\chi ={\cal O}\left(J^{-1}\right)$ at low
temperature .

Within DMFT \cite{dmft}, the dc resistivity is easily obtained
from one-particle properties since vertex corrections are absent.
In the Fermi-liquid regime $T<T^\star$, we find
$\rho_{dc}(T)\propto (T/T^{\star})^2$: hence the Kadowaki-Woods
relation is obeyed as in most heavy-fermion compounds.
 In the intermediate regime $T^{\star} < T < T_{
K}$, the resistivity drops as $\rho_{dc}(T)\propto
(T^{\star}/T)^{1/2}$. Although reminiscent of the maximum observed
in usual heavy-fermion systems the physical origin of this feature
in the present case is different. Here, the localized spins form
tightly bound singlet pairs in the SL phase at $T^{\star} \ll T \ll
J$. Hence,  the correlations between the localized
spins increase as
$T$ increases from $T^{\star}$, which results in the decrease of
 the scattering cross section of conduction
electrons. However, residual scattering of the
conduction electrons on the local spin fluctuations of the SL is
still expected to contribute to the resistivity for $T_K < T < J$.
The discussion of this effect requires going
 beyond the ${\cal N} \to \infty$ limit
(while we do not
expect qualitative changes in the FL region from the
inclusion of higher order corrections). There is in particular
a second-order
correction to the conduction electron self-energy that reads:
\begin{equation}
\label{correction} {\rm Im}\;\Sigma_c(0) \propto \left(J_K/{\cal
N}\right)^2\;\rho_0(0)\;\int_0^\infty\;\frac{d\omega}{\sinh\left(\beta
\omega\right)}\;\chi_{loc}''(\omega)\;.
\end{equation}
Since $\chi_{loc}''(\omega) \propto \omega/T$ for $\omega < T$ and $T > T^{\star}$, this results in a
contribution $\delta \rho \propto T$ to the resistivity. We thus
expect to see in the physical case ${\cal N} = 2$ a crossover from
a quadratic to a linear $T$-dependence of the resistivity at $T
\approx T^{\star}$.
\section{Conclusions}

In summary, we studied a model in which the competition between
the Kondo effect and frustrating magnetic interactions leads to a
QCP separating a heavy Fermi-liquid phase from a spin-liquid
phase. The coupling of the conduction electrons to critical local
spin-fluctuations near the QCP results in a dramatic reduction of
both the Kondo temperature and the Fermi-liquid coherence scale,
and to a critically enhanced effective mass. For temperatures
above the FL coherence scale but below the magnetic exchange, the
transport and spin dynamics show striking deviations 
from those of a Fermi liquid.

To conclude this paper, we would like to make several remarks regarding the 
possible physical relevance of this model, and of our findings. 

First, it is worth mentioning that some of the results derived here are
independent of the specific form of the kernel $\chi(\tau)$ but
results from the existence of an unstable fixed point that lies
between the Kondo phase and the spin fluctuation dominated phase
as pointed out in~\cite{se}. Some of our conclusions are therefore
expected to be valid beyond our specific model, as long as the 
time-decay of the local spin correlations is sufficiently slow.

Next, we would like to comment on the manner in which long-range 
magnetic order (LRO) can possibly affect our results. 
In the solution presented in this paper, spin-glass ordering does not appear because 
it is suppressed by the strong quantum fluctuations associated with 
the fermionic representation of spins in the large-${\cal N}$ limit \cite{gepasa}. 
Taking LRO into account in the present model thus requires either to consider 
bosonic representations \cite{gepasa} (which however makes the Kondo 
effect more difficult to treat technically) or to go beyond large-${\cal N}$  
in the fermionic case. It is clear that spin-glass ordering will show up 
at first order in the $1/{\cal N}$ expansion, in the mean-field (infinite 
connectivity) model. One can quantify this by noting that 
the criterion for the spin-glass transition reads \cite{gepasa}: 
$J\chi_{loc}(T,J) \propto \sqrt{{\cal N}}$. In our solution, 
$\chi_{loc}$ is proportional to $1/T^*$, which is given by (\ref{tstar}) 
near the QCP. Using this expression, we see that LRO should set in at a value of 
$J$ which is {\it smaller} than the critical value $J_c$ found above (at which the coherence scale 
vanishes): $J_{LRO}=J_c (1-c/{\cal N}^{1/4})$. Hence, we expect that, in high enough dimensions 
(where mean-field theory applies), 
the vanishing of the Kondo scale will be {\it preempted} by magnetic LRO. 
Nevertheless, some of the qualitative features found above may still be 
important in practice when the coherence scale is small enough at the transition 
point. 
The situation in low dimensions, when spatial fluctuations are stronger, 
is quite open. Whether it is possible to reach the ``coherence'' transition 
before LRO sets in, as suggested in Ref.\onlinecite{qimiao} is a fascinating  
question. 

Finally, we would like to address the qualitative relevance of some of the 
findings of this paper for the physics of 
LiV$_2$O$_4$ and other compounds such as YScMn$_2$ and
 $\beta$-Mn. LiV$_2$O$_4$ has a structure in
 which the V ions form a lattice of corner-sharing tetrahedra. This
 highly frustrated structure is known to lead to unconventional
 spin-liquid ground-states~\cite{ramirez} and recent NMR
 studies~\cite{loidl} indeed showed evidence of the presence of
 SL-like spin correlations in LiV$_2$O$_4$. This system is also
 close to a SG instability as it is known that
  small amounts of Zn doping on the
 Li sites results in SG freezing~\cite{fuyaue}. According to LDA + U
 band structure calculations~\cite{huri,band1} one of the three
 $t_{2g}$ V d-levels splits from the triplet in the crystal field
 of LiV$_2$O$_4$ and forms a highly correlated band. Due to the
 large Coulomb repulsion on the V site the corresponding Wannier
 states are singly-occupied thus playing a role similar to that of the 
 f-orbitals in conventional rare-earth based heavy-fermion
 systems. 

In Ref.\onlinecite{huri}, a Kondo lattice model was thus proposed to describe 
LiV$_2$O$_4$. This model in fact neglects the important {\it ferromagnetic}
on-site Hund's coupling between 
the itinerant and localised orbital~\cite{claudine}. Let us however 
comment on the estimate of the resulting coherence temperature made in 
Ref.~\onlinecite{huri}.   
The single-site Kondo temperature estimated there from the electronic 
structure calculations is $T_{ K}^{0}\approx
 550$K~\cite{huri}. This is an order of magnitude larger than
 the measured Fermi liquid coherence temperature $T^{\star}\approx
 25-40 $ K ~\cite{exp1}. Nozi{\`e}re's ''exhaustion''
 mechanism~\cite{pn} was invoked to explain this
 huge reduction of $T^{\star}$ with respect to $T_{ K}^0$. However,
  some doubts where recently cast~\cite{bugegr,nrg} on the validity
 of Nozi{\`e}res' estimate $T^{\star}  \propto  (T_{K}^{0})^2/D$
 ($D$ is the bandwidth)  and it has been suggested~\cite{bugegr}
 that $T^{\star} \propto T_{K}^0$ with a prefactor that is
 small only for very low values of the conduction-electron density,
 a situation that is not realized in LiV$_2$O$_4$. Hence, an
 alternative mechanism is needed in order to explain the observed
 reduction of $T^\star$. As shown in this paper, magnetic frustration does 
 provide such a mechanism in a Kondo lattice model. 

However, as pointed out in Ref.~\onlinecite{claudine}, 
 it is crucial to take into account the Hund's rule coupling 
in a realistic modelling of LiV$_2$O$_4$ and related compounds. 
Hence the results of the present paper cannot be straightforwardly 
applied to these compounds. 
However, we observe that the strong mass enhancment 
due to frustration has a simple {\it qualitative origin}: it is due to the 
storage of entropy at low-temperature associated with the spin-liquid state 
characteristic of frustrated systems. Hence, we expect that other 
models in which electrons are strongly coupled to local spins with 
SL dynamics will also lead to such a mass enhancment and suppression of 
the coherence scale. 
In LiV$_2$O$_4$, the $T$-independent spin relaxation rate for 
$T>T^\star$ as well as 
incoherent metallic transport $\rho\propto T$ in this regime is, in our 
opinion, strong experimental evidence that the local spin dynamics characteristic 
of a frustrated spin-liquid is playing a key role in this system.

\acknowledgments We thank C. Lacroix for useful comments and
discussions and N. Buettgen for making Ref.~\onlinecite{loidl} available to
us prior to
publication.
A.G acknowledges early discussions with A. Sengupta
and O. Parcollet on closely related
subjects as well as useful discussions with H. Takagi, V. Anisimov and 
M. Katsnelson on LiV$_2$O$_4$.


\begin{references}
\bibitem{footnote1} The explicit form of the last term in the 
electronic part of the action in Eq.~(\protect{\ref{action}})
 corresponds to the choice of a semi-circular conduction-band density of
states, $\rho_0(\epsilon) =2/\pi \sqrt{1 - \epsilon^2/D^2}$.
 The results that follow are qualitatively independent of
the detailed form of $\rho(\epsilon)$.
\end{references}
\end{document}